\documentstyle[epsf]{elsart}

\begin{document}
\begin{frontmatter}
\title{Study for the $s$-wave in $^{10}$Li with analysis of cross-sections}
\author[cor1]{Hiroshi Masui},
\author[cor2]{Shigeyoshi Aoyama},
\author[cor3]{Kiyoshi Kat\=o},
\author[cor4]{Kiyomi  Ikeda}
\address[cor1]{Meme Media Laboratory, Graduate School of Engineer,
        Hokkaido University,
        Sapporo 060-8628, Japan}
\address[cor2]{Information Processing Center, Kitami Institute of
        Technology, Kitami 090-8507, Japan}
\address[cor3]{Division of Physics, Graduate School of Science,
        Hokkaido University,
        Sapporo 060-0810, Japan}
\address[cor4]{The Institute of Physical and Chemical Research(RIKEN),
   Wako 351-0198, Japan}

\begin{abstract}
We study the effect of $s$-wave cross-sections
in $^{4}$He+$n$ and $^{9}$Li+$n$ elastic scattering reactions
by using the Jost function method (JFM).
In $^{5}$He ($^{4}$He+$n$), the $s$-wave pole of the $S$-matrix
does not contribute so much to the total cross-section.
On the other hand, in $^{10}$Li ($^{9}$Li+$n$),
the $s$-wave component can not be neglected
due to a relatively strong attraction for the $s$-waves
of the core+$n$ potential.
It is shown that the $^{9}$Li-$n$ potential,
which is  microscopically derived by taking into account
the  pairing-blocking effect for a $p$-wave
neutron, reproduces the $s$-wave pole close to the $^{9}$Li+$n$
threshold and strongly enhances the $s$-wave cross-section near
the threshold.
\end{abstract}
%
\end{frontmatter}
%
Study for unstable nuclei has great importance to 
understand the nuclear excitation and the structure
from a new point of view \cite{Ta96}.
It has been known that 
the unstable nuclei have some different features from stable nuclei.
For example, large root-mean-squared radii which are understood as
a ``halo'' structure, have been observed
for neutron-rich nuclei near the drip-line \cite{Ta85,Ta96}.
Among these nuclei, $^6$He and $^{11}$Li
are considered as the typical halo nuclei.
Theoretical approaches for these nuclei
have been done by using the core+$n$+$n$ model.
Binding energies and halo structures
have been studied in detail by using this model.
It is known well that, for $^{6}$He,
the core ($^{4}$He)+$n$+$n$ model can successfully reproduce
the essential nature of this halo nucleus.
However, 
for $^{11}$Li, to reproduce the binding energy,
it might necessary to introduce
some ``special'' mechanisms in the $^{9}$Li+$n$ subsystem,
e.g. a strong state dependence for even- and odd-parity states,
since the lack of the binding energy is too large.
Thompson and Zhukov proposed 
a relatively strong attraction for the $s$-wave in the $^{9}$Li-$n$ potential,
so that $^{10}$Li has an $s$-wave ``virtual'' state \cite{Th94}.
Using such potentials, they discussed that
the experimental binding energy of $^{11}$Li
($\sim 0.31$ MeV) is reproduced.
Despite of the success,
the mechanism of such a strong state dependence
was not clearly explained yet.
Recently, Kat\=o, Yamada and Ikeda proposed a potential with
a new mechanism that the $s$-orbit comes down to the same energy
region of the $p$-orbits
due to the pairing-blocking effect upon the pairing correlation
of $p$-orbital neutrons in the core \cite{Ka99}.
Using such a state dependent potential, 
we showed that the $s$-wave virtual state is obtained
in the $^{9}$Li+$n$ system (the scattering length: $a_{0} = -9.97$fm)
and the position of the pole is
$k = -i0.077$fm$^{-1}$ ($E_{v} = -0.135$MeV)
in our previous calculation \cite{Ma00}.

From the experimental side,
it remains to discuss whether the $s$-state of $^{10}$Li is
a resonant state \cite{Yo94,Zi97,Sh97}
or a virtual one \cite{Zi95,Sh98,Th99}.
In the resonant state cases, the energies are observed
within $0.1 \sim 0.2$ MeV and the widths are $\sim 0.5$ MeV.
On the other hand, in the virtual state cases,
where the state is characterized as the scattering length $a_{0}$,
the estimated scattering lengths are $a_{0} < -20$ fm. 
The effect of a resonance pole is observed as a ``peak''
in the total cross-section,
and poles of bound and virtual states make an enhancement
at the core+$n$ threshold energy.
The general behavior of the pole near the threshold energy
and its effect to the total cross-section
have been investigated \cite{Fr64}.
In the $^{9}$Li+$n$ system,
it is suggested that there exists an $s$-state
near the threshold energy despite of no bound state,
from both experimental \cite{Zi95,Sh98,Th99}
and theoretical sides \cite{Th94,Ma00}.
Hence, it is worthwhile to investigate the
behavior of the pole and effect to the cross-section in the
$^9$Li+$n$ system.

In this paper,
we calculate the total cross-sections
of the $^{4}$He+$n$ and $^{9}$Li+$n$ elastic scattering
and investigate the contribution of the $s$-waves in the cross-sections.
And discuss the effect of the $s$-wave virtual state in $^{10}$Li
with comparing an experimental result for a measurement of
energy distributions of $^{9}$Li and a neutron
in projectile fragments:$^{11}$Li+C
$\rightarrow$ $^{9}$Li+$n$+X \cite{Ko93}.

First, we explain the method for the numerical calculation.
To study cross-sections and 
the effect of the $s$-wave poles,
we use the Jost function method (JFM)\cite{So97,Ra98},
since JFM is quite useful to obtain the position of poles both
in the complex energy and the momentum planes.
Details of the procedure for an actual calculation are given in 
the original papers of Sofianos and Rakityansky \cite{So97,Ra98}
and also in our previous papers \cite{Ma99,Ma00}.
The important features of the JFM are follows:
First, introducing unknown functions ${\cal F}^{(\pm)}(k,r)$
for coefficients of the incoming $(-)$ and outgoing $(+)$ waves,
we express the regular solutions of the Schr\"odinger equation as
\begin{equation}
\phi(r) \equiv
\frac{1}{2}
\left[
  H^{(+)}_{l}(k r)  {\cal F}^{(+)}(k, r)
+ H^{(-)}_{l}(k r)  {\cal F}^{(-)}(k, r)
\right]
\label{eq:phi_to_h_pm}
\mbox{ .}
\end{equation}
Second, to solve the equation for ${\cal F}^{(\pm)}(k,r)$,
we introduce the additional constraint condition,
which is usually chosen in the variable-constant method:
\begin{equation}
H^{(+)}_{l}
\left[
  \partial_r {\cal F}^{(+)}
\right]
  + H^{(-)}_{l}
\left[
  \partial_r {\cal F}^{(-)}
\right]
=0
\mbox{ .}
\label{eq:h_const}
\end{equation}
Using this condition, the equation for ${\cal F}^{(\pm)}(k,r)$
becomes a first-order one:
\begin{equation}
\frac{\partial {\cal F}^{(\pm)}(k,r)}{\partial r}=
\pm \frac{\mu}{i k\hbar^{2}}
H^{(\mp)}_{l}
V(r)
\left\{
    H^{(+)}_{l}{\cal F}^{(+)}(k,r)
  + H^{(-)}_{l}{\cal F}^{(-)}(k,r)
\right\}
\mbox{ .}
\label{eq:Jost_f_dif}
\end{equation}
In the asymptotic region,
where the potential goes zero faster than the Coulomb force,
the ${\cal F}^{(\pm)} (k,r)$ converge into constant ${\cal F}^{(\pm)}(k)$,
and are equivalent to the ``Jost functions'' of the original
Schr\"odinger equation.
By using the Jost functions ${\cal F}^{(\pm)}(k)$,
the $S$-matrix is expressed as
\begin{equation}
  S (k)=
  {\cal F}^{(+)} (k) / 
 {\cal F}^{(-)}(k)
   \mbox{ .}
   \label{S-matrix_Jost}
\end{equation}
With the help of definite shape of the asymptotic wave function,
the convergence of the Jost functions ${\cal F}^{(\pm)}(k)$ is quite well.
Hence, by using JFM, the physical quantities
concerned with the $S$-matrix can be obtained very easily and accurately.
Moreover, we applied this method 
to obtain the $s$-wave poles of the $S$-matrix for virtual states
in the complex-momentum plane
by solving ${\cal F}^{(-)}(k) = 0$ \cite{Ma00}.

Next, we show the results
of numerical calculations for cross-sections
of the elastic scattering.
For the $^{4}$He+$n$ scattering,
we use the KKNN potential \cite{Ka79}.
Although the $^{4}$He+$n$ elastic cross-section
was already given in Ref.\cite{Ka79},
we re-calculate it to emphasize a less contribution of the $s$-wave
component to the total cross-section.
In Fig.~1, it is shown that the dominant component of the
total cross-section (solid-line) is the $p_{3/2}$ component (dotted-line)
and the $s_{1/2}$ component (dashed-line)
has actually a less contribution.
The result is reasonable, since the $S$-matrix pole of
the $s$-wave resonance in the $^{4}$He+$n$ system using the KKNN potential
has a very large imaginary part; $E  = 7.50 - i44.7$MeV,
as shown in our previous paper \cite{Ma00}.
Hence, the contribution of the $s$-wave
to the total cross-section is very small.
In Fig.~1, the peak around 1 MeV in laboratory frame
corresponds to the $p_{3/2}$ resonance 
($E_r=0.89 \pm 0.05$ MeV, $\Gamma=0.60\pm0.02$ MeV \cite{Aj88},
in the center of mass energy),
and our calculational results($E_r=0.74$ MeV and $\Gamma=0.59$ MeV)
using the KKNN potential almost agrees with it.
The shape of the total cross-section
and the position of the peak corresponds to
the observed cross-section (see Fig.~1).

\begin{center}
------------\\
Fig.1\\
------------\\
\end{center}

For the  $^{9}$Li+$n$ scattering,
we use the same type of the folding potential
applied in our previous calculation \cite{Ka99,Ma00}.
To see the contribution of each partial wave clearly,
we perform a single-channel calculation
using a parameterization of Ref.\cite{Ka99} for $p$-waves,
and investigate the difference by changing
the parameter for $s$- and $d$-waves.
In this calculation, we assume that the spin of the core nucleus
($^{9}$Li) is $\frac{3}{2}^{-}$
coming from the valence $p_{3/2}$ proton,
and we assign the quantum numbers
as $(l_{j} \,; J ^{\pi})$ for the core+$n$ elastic scattering,
$[p_{3/2} (^{9}\mbox{Li}) \otimes l_{j}(n)]_{J^{\pi}}$.

In Fig.~2(a), we show the total cross-section of
the $^{9}$Li+$n$ scattering.
In our previous paper \cite{Ma00},
the 1$s$-state in the $^{9}$Li+$n$ system is obtained as the
``virtual'' state ($a_{0} = -9.97$fm),
which has the pole on the complex momentum
plane as $ k = - i c \,\,(c > 0)$.
A possibility for the existence of the $s$-wave virtual state
in the $^{9}$Li+$n$ system is
also suggested by Thompson and Zhukov \cite{Th94}.
Due to the contribution of the virtual $s$-wave pole,
$k = -i0.077$ fm$^{-1}$ ($E_{v} = -0.135$ MeV) for ($s_{1/2}\, ; 2^{-}$),
the total cross-section has an enhancement at the threshold energy.
The contribution of each partial wave is also shown in
the same figure (Fig.~2(a)). 
The sharp peak at 0.5 MeV of the total cross-section comes from
the contribution of the resonance poles of the $p$-waves
($p_{1/2}\, ; 1^{+}$) and ($p_{1/2}\, ; 2^{+}$)
at $E = 0.422 - i 0.161$ MeV and $0.542 -i 0.242$ MeV, respectively,
using this potential.
And for the $s$-waves,
the position of the pole for the ($s_{1/2}\, ; 2^{-}$)
is closer to the threshold energy than the ($s_{1/2}\, ; 1^{-}$)
one at $k = -i0.127 $ fm$^{-1}$ ($E=  -0.366$ MeV).
Therefore, we investigate the effect of the ($s_{1/2}\, ; 2^{-}$) pole
to the total cross-section.
To see the contribution of the $s$-wave pole clearly,
we change the strength of the attractive part
of the $^{9}$Li-$n$ potential \cite{Ka99} (see Fig.~2(b)).
The potential is made by taking into account the dynamics of
a pairing-blocking effect of $p$-wave neutrons for the core nucleus
and a pairing correlation mechanism,
then the potential depth for $p$-waves is pushed up
and the attraction of the $s$- and the $d$-wave (even-parity states)
potential becomes relatively stronger than the $p$-wave.
In this model, the strength
for the $s$- and $d$-waves still has some freedom to fix parameters.
Hence, we slightly vary the strength for the even-parity state
and investigate the relation between the position of the $s$-wave poles
and the total cross-sections.
In our previous calculation, which is one of the candidates,
the $s$-wave pole is at $k = -i0.077$ fm$^{-1}$ on the complex momentum
plane (virtual state) for ($s_{1/2}\, ; 2^{-}$).
Then, we have done the calculation with two different strengths,
one is more attractive,
which is determined so as to have the pole at almost origin;
virtual state at $k = -i0.023 $ fm$^{-1}$  ($E = -0.012$ MeV)
and another one is less attractive,
which is the same order of the strength for the $p$-wave interaction;
virtual state at $k = -i0.36 $ fm$^{-1}$ ($E = -2.89$ MeV).

\begin{center}
------------\\
Fig.2(a)(b)\\
------------\\
\end{center}

Each calculated total cross-section shows
a different behavior at the threshold energy.
For the weak strength,
there is no enhancement,
since the $s$-wave pole is very far from the threshold energy,
$E = -2.89$MeV ($k = -i0.36 $ fm$^{-1}$).
Hence, the contribution of the $s$-waves to the total cross-section
using this strength becomes very small.
Also, for the normal case,
the enhancement is not so large,
due to the small contribution of the $s$-waves cross-sections
as shown in Fig.~2(a).
On the other hand, for the strong case,
the enhancement becomes very large,
since the position of the $s$-wave pole is
in close proximity to the threshold energy,
$E = -0.012$ MeV ($k = -i0.023 $ fm$^{-1}$),
and the contribution to the total cross-section also becomes 
large at the threshold.

For the measurement of the states of $^{10}$Li,
charge-exchange reactions such as
$^{9}$Be($^{13}$C,$^{12}$N)$^{10}$Li \cite{Bo93}, etc.,
proton stripping reaction
such as $^{11}$Be($^{7}$Li,$^{8}$B)$^{10}$Li \cite{Yo94},
pion-absorption such
as $\pi^{-}$+$^{11}$Be $\rightarrow$ $p$+$^{10}$Li \cite{Am90},
and projectile fragments
as $^{11}$Li+C $\rightarrow$ $^{9}$Li+$n$+X \cite{Ko93}
have been used.
Due to the difficulty of experiments,
there are no available data for direct measurements of
reactions between a neutron and $^{9}$Li in contrast
to the $^{4}$He+$n$ case,
and it would be one of the most challenging subjects
in study of the unstable nuclei.
Among the above experiments,
we consider that the energy distribution of the projectile fragments
$^{9}$Li+$n$ in $^{11}$Li+C $\rightarrow$ $^{9}$Li+$n$+X reaction
has some correspondence to the total cross-section.

In Ref.\cite{Ko93},
the author used unstable nuclei beams ($^{6}$He and $^{11}$Li)
bombarding to a carbon target,
and measured the energy distribution of the projectile fragmentation:
the $^{4}$He+$n$ in  $^{6}$He+C$\rightarrow ^{4}$He+$n$+X, and
the $ ^{9}$Li+$n$ in $^{11}$Li+C$\rightarrow ^{9}$Li+$n$+X.
In these reactions, it was assumed that
a sequential decay process is dominant \cite{Ko93}.
In the first step of these reactions,
one valence neutron is removed from the projectile,
and the remainder forms an intermediate resonance.
In the second step,
the resonant nucleus emits a second neutron.
The energies of the second neutron;$a$ and the rest particle;$b$
are measured, and the relative energy between these particles
are determined by using the invariant mass method,
\begin{equation}
  \label{eq:inv_mass}
  E_{\mbox{\tiny c.m.}} = \sqrt{(E_{a} + E_{b})^{2}
    -(p_{a} + p_{b})^{2}} -m_{a} - m_{b}
\mbox{ .}
\end{equation}

When the sequential decay picture is the dominant process
in the reaction,
we can consider that
an energy distribution
gives information of the intermediate resonant nucleus
($^{5}$He or $^{10}$Li).
Here, we consider the $^{10}$Li case.
In the projectile fragmentation,
the intermediate ``resonant'' nucleus $^{10}$Li,
decays apart into a neutron and a core nucleus $^{9}$Li \cite{Ko93}.
Then, at the resonant energy,
the energy distribution of the projectile fragments
has a resonant peak due to the contribution of the intermediate
resonant particle $^{10}$Li.
Also, it is enhanced at the threshold energy when
the ``virtual'' state exist in the intermediate state.
On the other hand,
it is well known that the elastic cross-section has a peak at the
resonant energy, and is enhanced at the threshold energy
when a virtual state exists.
Therefore, the energy distribution of the fragmentation
for the low-energy is considered to
be similar to ones of the elastic scattering,
since the elastic cross-section also has the peak for the resonant state
and the enhancement for the virtual state.

In the upper part of Figs.~3(a) and ~3(b),
we show our calculated results
for the total cross-sections.
To compare the calculated total cross-sections with the
energy distribution of the experiment \cite{Ko93},
we smear the cross-sections using the resolution of the detector, $\Gamma$.
And in the lower part,
we show energy distributions of $^{4}$He+$n$ ($^{5}$He)
and $^{9}$Li+$n$ ($^{9}$Li)
in the experiment of projectile fragments 
obtained in Ref.\cite{Ko93}.

\begin{center}
------------\\
Fig.3(a)(b)\\
------------\\
\end{center}

In the $^{5}$He case,
the correspondence between 
the calculated total cross-section
and the energy distribution is good by taking into
the resolution of the detector as $\Gamma = 2.2 \sqrt{E}$ MeV \cite{Ko93},
(see Fig.~3(a)).
Based on the good correspondence with the $^{5}$He case,
we investigate the $^{10}$Li case as shown in Fig.~3(b).
Our calculation of the total cross-section shows
that the enhancement at the threshold energy
comes from the $s$-wave virtual state,
as discussed in Fig.~2(a).
We obtain a good correspondence
by using $\Gamma = 0.45 \sqrt{E}$ MeV \cite{Ko93}.
Therefore, we can make sure that
the enhancement at the threshold energy comes
from the effect of the $s$-wave virtual state.
The closer the position of the pole comes to the threshold energy,
the higher the enhancement goes,
as shown in Fig.~2(b).
Suppose that the $^{10}$Li has no sharp $s$-wave resonance
at low energy  ($ < 1$ MeV), such as our case,
the enhancement only comes from the $s$-wave virtual state.
From the experimental side,
there are several measurements
that indicate the $s$-wave virtual state \cite{Zi95,Sh98,Th99}.
The virtual state is characterized
by using the scattering length $a_{0}$ as discussed above.
In Ref.\cite{Th99},
the authors estimated the scattering length
as $a_{0} = -16^{+4}_{-7}$ fm
by using the effective range theory.
This result corresponds our results
within a strong attractive case ($a_{0} = -40.8$fm)
and original one ($a_{0} = -9.97$fm).
Our potential model, which is made by microscopically
with taking into account a pairing-blocking mechanism,
has no sharp $s$-wave resonance at low energy ($ < 1$ MeV)
with varying the strength of the attraction \cite{Ma00},
Hence, at this stage,
the enhancement of the energy distribution at the threshold energy 
can be considered as an effect of the $s$-wave virtual state.

In summary, we calculate the total cross-sections
of the $^{4}$He+$n$ and $^{9}$Li+$n$ elastic scattering.
As we expected, we found a difference in the contribution
of the $s$-wave between the $^{4}$He+$n$ ($^{5}$He)
and the $^{9}$Li+$n$ ($^{10}$Li) systems.
In the $^{9}$Li+$n$ system,
the effect of the $s$-wave virtual pole appears as
the enhancement at the threshold energy.
we obtained a good correspondence between 
the calculated total cross-section
and the observed energy distribution \cite{Ko93}.
From the comparison of the process of them,
we can consider that these two quantities are related to each other.
Hence, the character of the resonance peak
and the enhancement at the threshold energy
corresponds between the cross-section and the energy distribution. 
Therefore, we can conclude, at this stage, that
using our potential model, there is no low-lying
$s$-wave resonant state but is a ``virtual'' state,
and then the enhancement of the energy distribution shows 
the existence of the $s$-wave virtual state
with large possibility.

In the future work,
we should perform coupled-channel calculations of the cross-sections
to investigate the property of $^{9}$Li-$n$ reaction precisely,
and must proceed to study the reaction mechanism of
$^{11}$Li+C$\rightarrow ^{9}$Li+$n$+X process.

\section*{Acknowledgments}
One of the authors (H. Masui) thanks Professor Y. Tanaka (VBL) for
giving a chance for a post-doctoral fellowship.
The authors are grateful to Dr. Ohnishi for a fruitful discussion
and also, the laboratory members of nuclear theory group
at Hokkaido University for helpful discussions. 


\newpage
\begin{figure}
  \epsfxsize=15cm \centerline{\epsffile{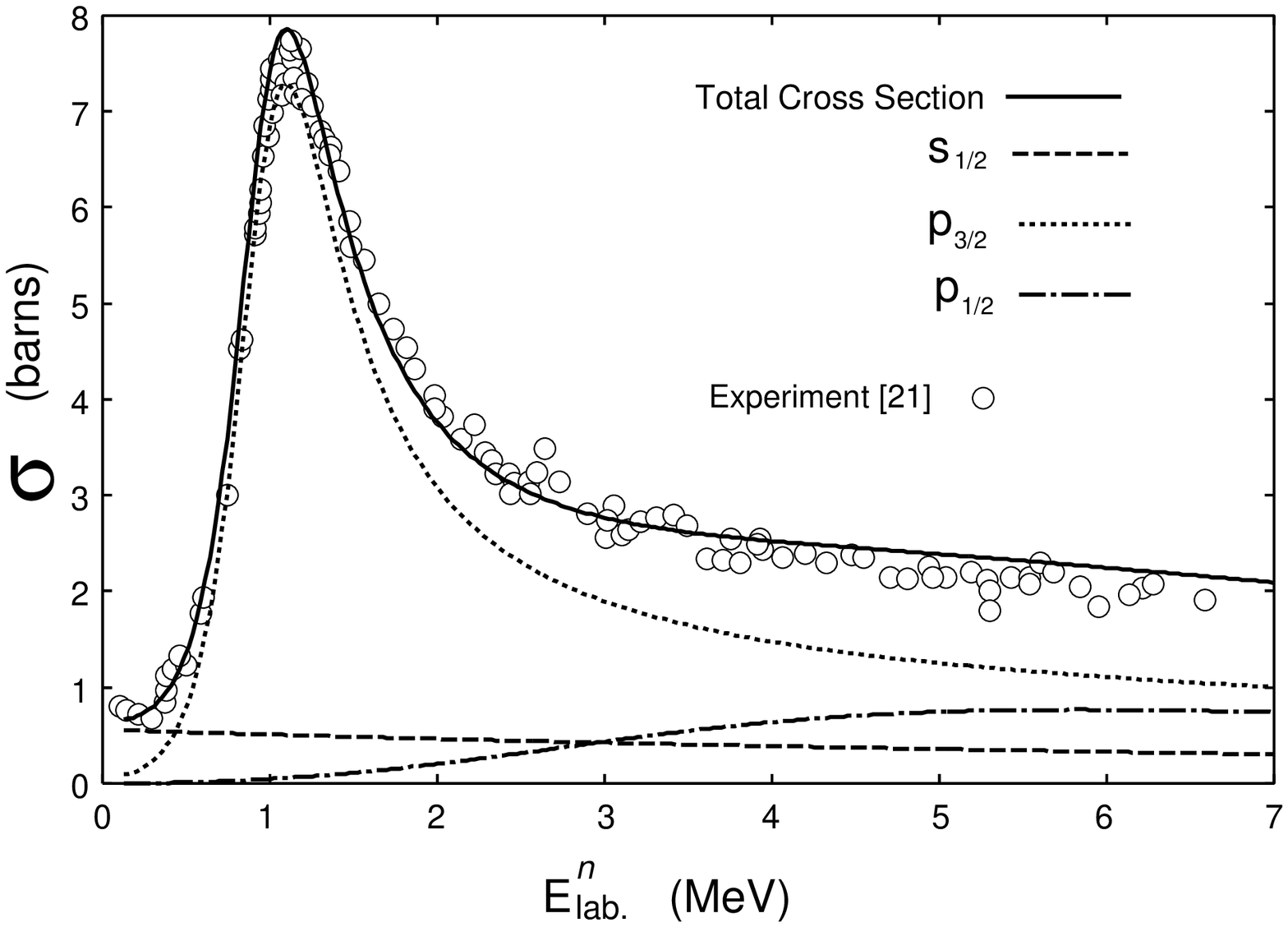}}

  \caption{Total cross-sections of the $^{4}$He+$n$ elastic scattering.
    The energy is taken as the laboratory frame.
    Solid line is the sum of the angular momenta (up to $l =3$).
    Dashed line is the contribution of the $s_{1/2}$ wave.
    Dotted line and dash-dotted line are $p_{3/2}$ and
    $p_{1/2}$ wave, respectively.
    The open circles are experimental data \protect\cite{Mo68}}

\end{figure}
\newpage
\begin{figure}
\epsfxsize=15cm \centerline{\epsffile{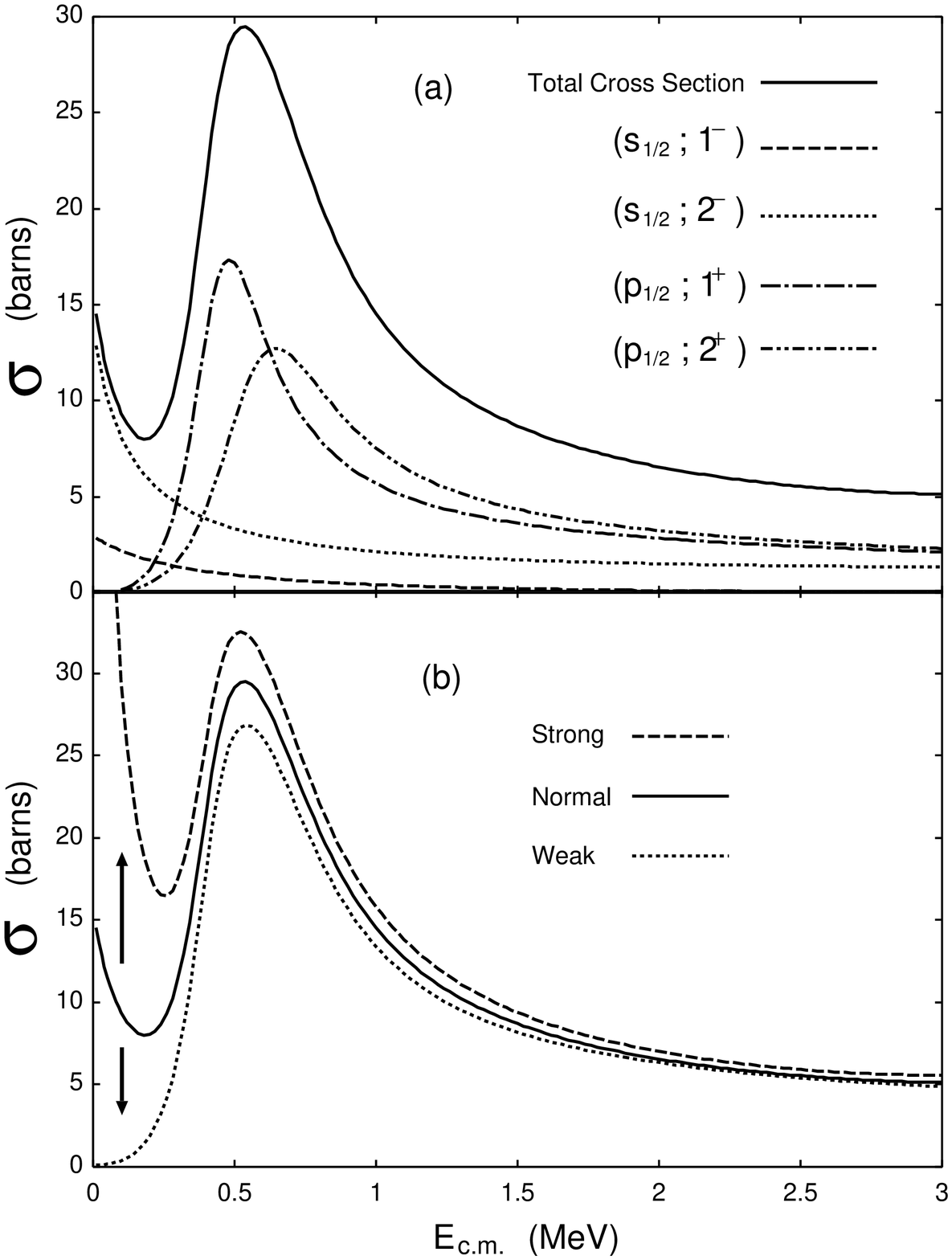}}

\caption{(a)Total cross-sections of the $^{9}$Li+$n$ elastic scattering.
  The energy is taken as the center of mass system.
  Solid line is the sum of all angular momenta.
  Others are contributions of each partial wave assigned as
  ($l_{j} \, ; J^{\pi} $).
  (b)The effect of the $s$-wave virtual pole on the total cross-section.
  Solid line is the total cross-section for normal strength,
  which has the virtual $s$-wave pole
  at $E= -0.135$ MeV same as our previous calculation \protect\cite{Ma00}.
  Dashed line is strong one and dotted line is weak one in
  the attractive part of the $^9$Li-$n$ potential.}

\end{figure}
\newpage
\begin{figure}
  \epsfxsize=17cm \centerline{\epsffile{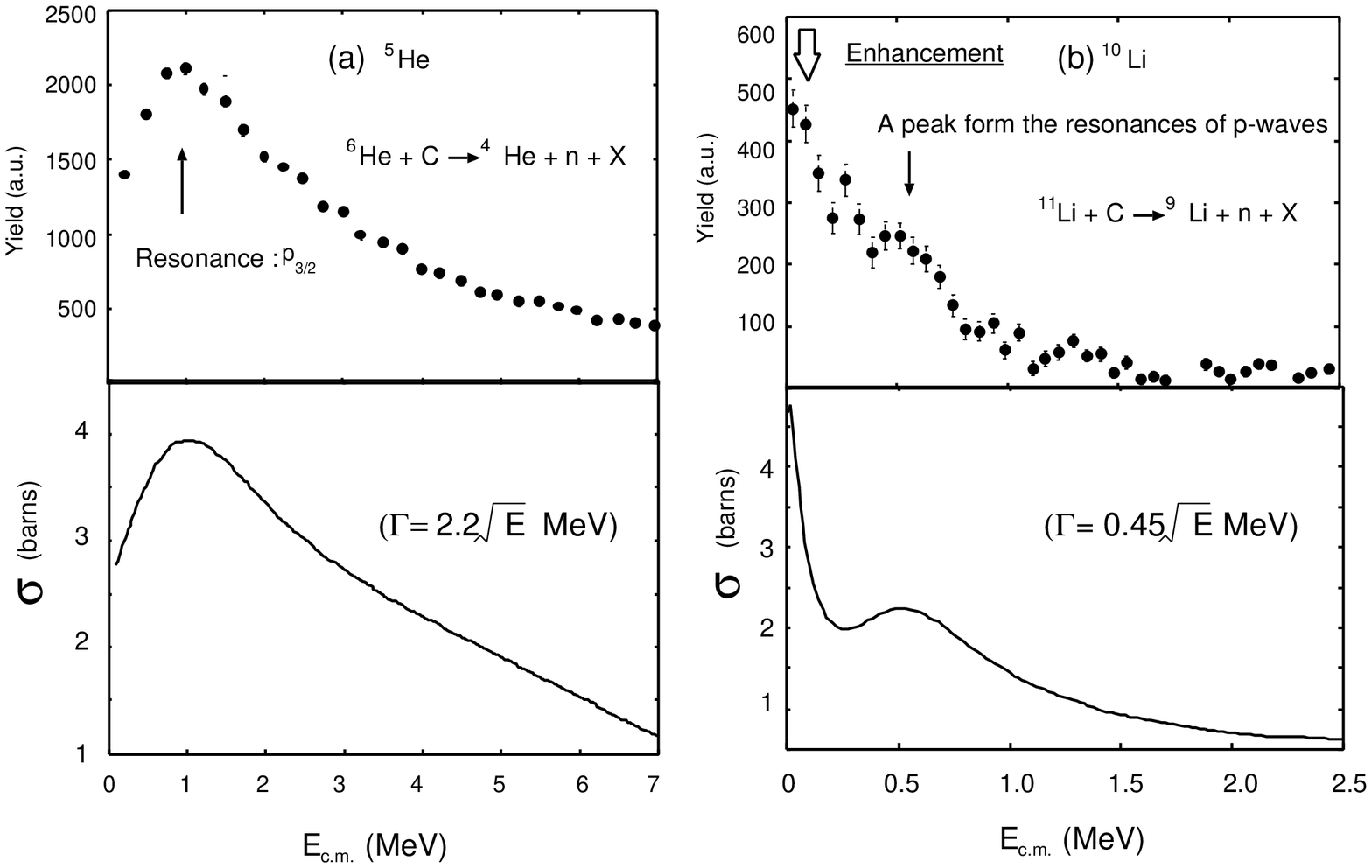}}

  \caption{Upper part:
    (a) the total cross-section of
    the $^{4}$He+$n$ elastic reaction,
    after taking into account the resolution of the detector
    as $\Gamma = 2.2 \sqrt{E}$ MeV and
    (b) the $^{9}$Li+$n$ case as $\Gamma = 0.45 \sqrt{E}$ MeV.
    Lower part:
    (a) the energy distributions of the $^{4}$He+$n$ ($^{5}$He)
    system from $^{6}$He+C at 800 MeV/A
    and (b) the $^{9}$Li+$n$ ($^{10}$Li) system from
    $^{11}$Li+C at 72 MeV/A, which are obtained
    by the experiment of Ref.\protect\cite{Ko93}.
    }
\end{figure}
\end{document}